\documentclass[twocolumn,prl,superscriptaddress,showpacs,preprintnumbers,amsmath,amssymb]{revtex4}

\usepackage{bm}
\usepackage{slashbox}
\usepackage{graphicx}
\usepackage{dcolumn}
\usepackage{amsmath}
\usepackage{amssymb}

\begin{document}

\title{Ionic mixtures in two dimensions: from crystals to chain and dipole gases}

\author{Lahcen Assoud, Ren\'e Messina, Hartmut L\"owen}
\affiliation{
 Institut f\"ur Theoretische Physik II: Weiche Materie,
 Heinrich-Heine-Universit\"at D\"{u}sseldorf,
 Universit\"{a}tsstrasse 1,
 D-40225 D\"{u}sseldorf, Germany }


\begin{abstract}
The ground state of a two-dimensional ionic mixture
composed of oppositely charged spheres is determined as a function of the 
size asymmetry by using a penalty method. 
The cascade of stable structures  includes square, triangular
and rhombic crystals as well as a dipolar pair gas and a 
gas of one-dimensional crystalline chains.
Thereby we confirm the square structure, found experimentally on charged
granulates, and predict new phases detectable in experiments on granular
and colloidal matter.
\end{abstract}


%
%

\maketitle

Predicting the crystalline structures from first principles 
is one of the  key problems in condensed 
matter physics, material science, chemistry, geophysics and polymer physics
\cite{Catlow}. For three-dimensional ionic crystals \cite {Madelung,Ewald},
it is common textbook knowledge \cite{Evans} that there are three basic equimolar 
structures whose stability depends on the ratio of the ion radii.
For increasing asymmetry in the radii, the corresponding stability sequence involves the 
cesium-chloride, sodium-chloride, and zincblende structures. As indicated by the names, 
these structures are realized in nature for the (ionic) salt crystals $CsCl$, 
$NaCl$ and $ZnS$ respectively, but they show also up for a number of other equal-valency salt
crystals and for mesoscopic oppositely charged colloidal particles which are
suspended in a strongly deionized solvent \cite{Leunissen,Dijkstra,Dijkstra2}. The
stability is rationalized by minimizing the Madelung  potential energy per particle pair
with respect to various periodic equimolar candidate lattices at zero temperature and zero pressure.

In two spatial dimensions, a similar problem arises by exploring
the ground-state structure of ionic crystal monolayers in a model of oppositely charged disks
with different radii. In fact, three different realizations of two-dimensional ionic 
crystals are possible on microscopic, mesoscopic and macroscopic length scales. First of all,
 crystalline sheets  of molecular salts can be deposited on smooth substrates 
by using e.g. electrochemical methods \cite{Rosendahl}. 
Secondly, oppositely charged colloidal particles  \cite{Leunissen,Dijkstra} can be confined 
to a single layer, e.g. by
using laser-optical traps \cite{traps} or suspending them at a flat 
fluid-fluid interface \cite{interface_Goedel}. Finally, there are  granulate systems
of millimeter-sized metallic balls which are oppositely charged and self-organize 
on a macroscopic plate into crystalline arrays \cite{Whitesides1,Whitesides2}. 
It is important to understand the different crystalline sub-structures 
both from a fundamental point of view and for applications.
For example, a control of the  composite structures
of colloidal crystals leads to new photonic \cite{Pine} or phononic \cite{Bechinger_PRL}
band-gap materials,
to molecular-sieves \cite{Kecht}, to micro- and nano-filters 
with desired porosity \cite{Goedel1} and to nanowires composed of individual particle strings
\cite{nanowires}. It allows to steer protein crystallization \cite{Nagayama}.

Despite its fundamental importance, the stability of two-dimensional ionic monolayers
has not been addressed for asymmetrically sized ions. Previous theoretical studies have mainly 
focused on ionic criticality \cite{Fisher_Levin} at finite temperature and on equal ion size \cite{Caillol_1998}.
In this Letter, we predict the ground state structures for two different relevant set-ups of oppositely
charged  spheres: in the first, all centers of mass belong to a common plane which
corresponds to a situation where both species are confined in an interface
\cite{Goedel1}. In the second set-up,
all particle surfaces touch a common plane corresponding to spheres in contact with the same planar substrate 
where they are confined by e.g.\ gravity \cite{Whitesides1,Whitesides2}. 
Using a novel penalty method, a wealth
of different stable structures is found. These include periodic crystalline arrays, 
on the one hand, with square, triangular
and rhombic unit cells and with both touching and non-touching large spheres.
On the other hand,  a dipolar gas of particle pairs and a gas of particle chains
with three different internal chain structures are also stable for strong size asymmetries
in the substrate set-up. 
All these structures are detectable in experiments on granular
and colloidal matter and in adsorbed crystalline layers of molecular salts.

The model system used in our study are equimolar $1:1$ mixtures of
large ions (component $A$) with radius $R_A$ carrying a positive charge  
$q_A$ and small ions (component $B$) with radius $R_B$  carrying a  negative 
charge $q_B$ ($|q_A|=|q_B|\equiv q$) such that the whole system is electroneutral. 
The size asymmetry  $0\leq R_B/R_A \leq 1$ is 
denoted with $\sigma$.
These constitutive ions interact via a pairwise potential  
composed by a Coulombic and hard-core part
\begin{equation}\label{eq1}
u_{\alpha\beta}(r)=
\begin{cases}\frac{q_{\alpha}q_{\beta}}{r} & {\rm if}\; r \geq R_{\alpha} + R_{\beta}
\qquad (\alpha,\beta\,=\,A\;{\rm or }\;B)\\
\infty & {\rm  if}\; r < R_{\alpha} + R_{\beta},
\end{cases} 
\end{equation}
where $r$ is the center-center distance between the ions $\alpha$ and $\beta$.
The stability of the crystalline structure (at zero temperature) 
is ensured by steric interactions of the hardcore form.
Our objective is to determine the stable structures at
zero pressure and zero temperature  by numerical minimization
of the total potential energy.

Traditional minimization schemes \cite{GA,PRE_Dijkstra_2009} typically require a continuous
pairwise potential form in order to localize the minimum. The discontinuous hard-core potential
splits the parameter space into various distinct regions which hampers 
straight-forward numerical minimization. For potentials which involve a hard-core part
we have developed here a new technique to overcome this difficulty that relies 
on the so-called {\it penalty} method \cite{Fiacco} which was hitherto 
applied to geometric packing problems of hard bodies. The key idea is to relax
the non-overlap condition (i.e., $r \geq R_{\alpha} + R_{\beta}$) by introducing
an auxiliary penetrable pair interaction:
\begin{equation}\label{eq2}
v_{\alpha\beta}(r)=\frac{q_{\alpha}q_{\beta}}{r}
+\mu\;{\rm max}(R_{\alpha}+R_{\beta}-r,0), 
\end{equation}
where $\mu>0$ is a penalty parameter which is larger than the gradient of the potential at contact.
This parameters gives a finite energy penalty to any overlapping configuration.
If $\mu$ is finite but sufficiently large, 
the total potential energy for the auxiliary potential $v_{\alpha\beta}(r)$
has exactly the same minimal configuration as that for $u_{\alpha\beta}(r)$.
However, numerically the potential energy landscape is now continuous such that standard
minimization routines like the  simplex algorithm \cite{simplex} can be applied.
Though the penalty technique itself is applicable to any spatial dimensionality and hard particle shape,
we exploit it here to predict the ground state for binary hard charged spheres.

In detail, we consider a parallelogram as a primitive cell 
which contains $n_A$ $A$-particles and $n_B$ $B$-particles.
We restrict ourselves to the case $n_A<3,n_B<3$.
This cell can be described geometrically by the two spanning lattice vectors ${\mathbf a}$
and ${\mathbf b}$.
The position of a particle $i$ (of species $A$)
and  that of a particle $j$ (of species $B$) in the parallelogram is specified by the vectors
${\mathbf r}_{\rm i}^A=(x_i^{A},y_i^{A})$ and
${\mathbf r}_{\rm j}^B=(x_j^{B},y_j^{B})$, respectively. 
Thereby, the new potential energy function that needs to be minimized at zero pressure and zero
temperature as a function of the crystalline lattice parameters reads
\begin{eqnarray}\label{eq_energy}
\lefteqn{U_{\rm total}=\frac{1}{2}\sum_{\alpha=A,B}
\sum_{i,j=1}^{n_\alpha}\sideset{}{'}\sum_{\mathbf{L}}v_{\alpha\alpha}
(\mathbf{r}^\alpha_i-\mathbf{r}^\alpha_j+\mathbf{L})}\nonumber\\
& &+\sum_{i=1}^{n_A}\sum_{j=1}^{n_B}\sum_{{\mathbf L}}v_{AB}(\mathbf{r}^A_i-\mathbf{r}^B_j+\mathbf{L}),
\end{eqnarray}
with ${\mathbf L}=k{\mathbf a}+l{\mathbf b}$ where $k$ and $l$ are integers.
The sums over  ${\mathbf L}$ in Eqn.\ \ref{eq_energy}  run over all lattice cells where the prime indicates
that for  ${\mathbf L=0}$ the terms with $i=j$ are to be omitted.
In order to handle efficiently the long range nature of the Coulomb interaction, we employed
a Lekner-summation \cite{brodka_lekner}.





\begin{figure}
\begin{center}
     \includegraphics[width=8cm]{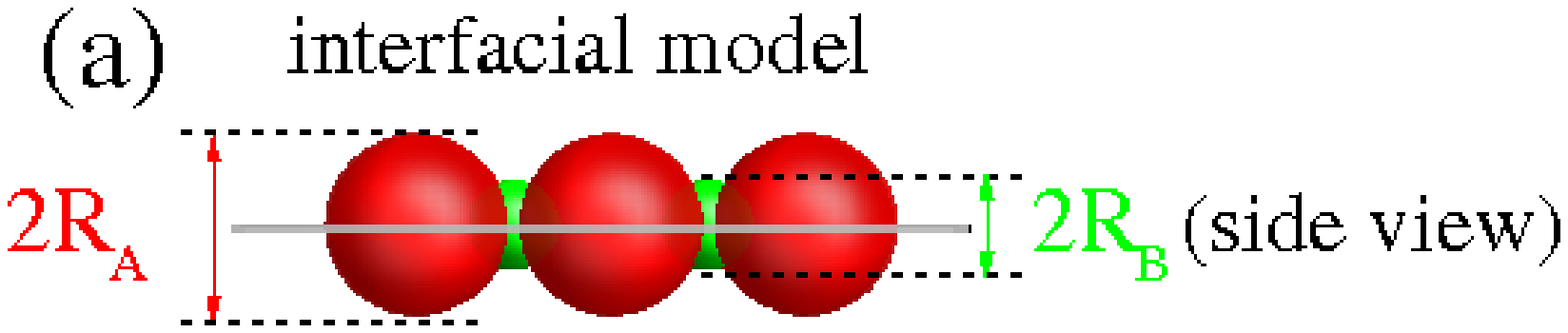}
     \includegraphics[width=8cm]{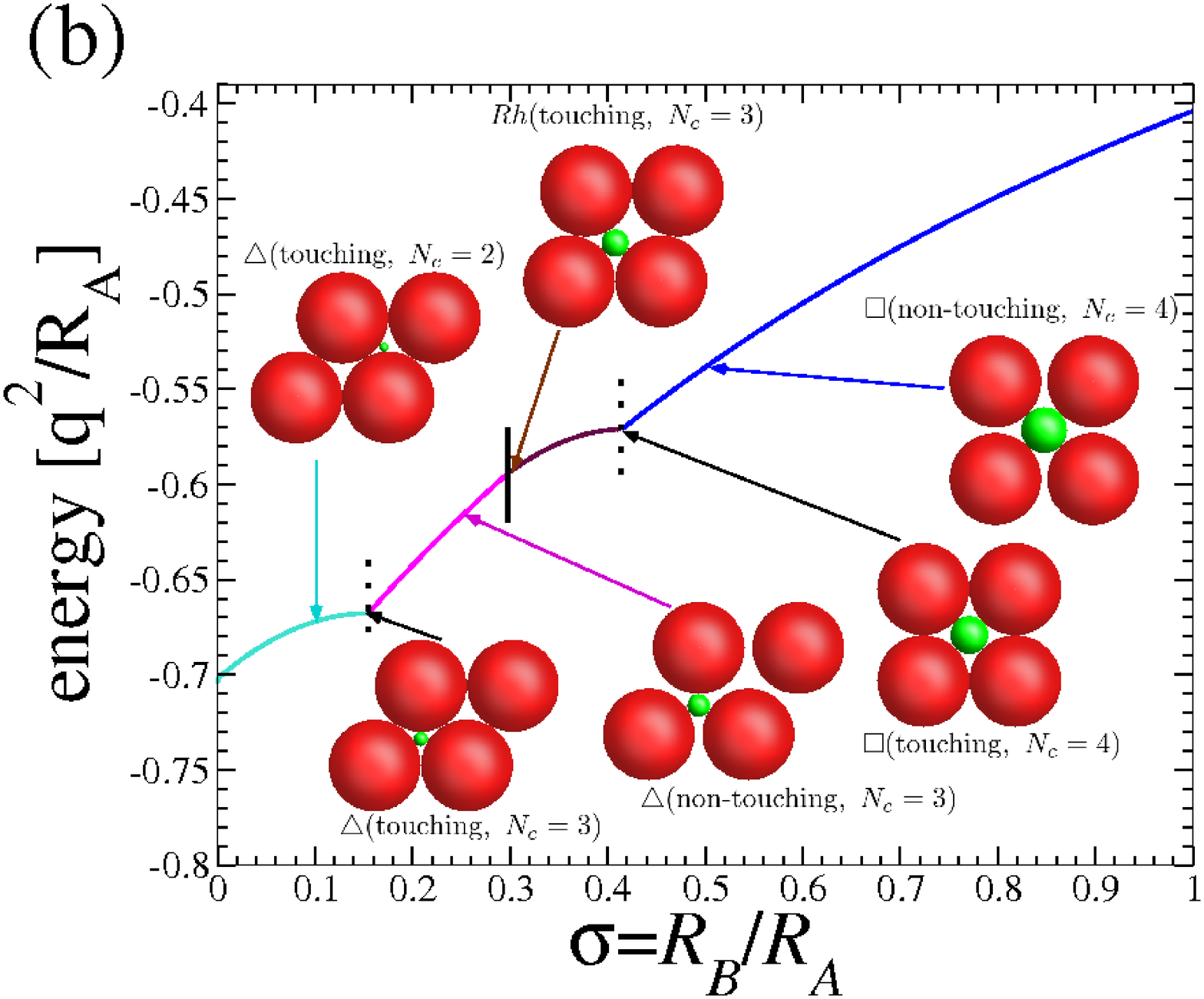}
\end{center}
\caption{Stable structures of oppositely charged spheres 
versus their size asymmetry $\sigma=R_B/R_A$ in the interface model,
where all sphere centers fall on the same plane: a)  side-view,
b)  (scaled) energy per ion. The discontinuous transition 
is indicated by a solid bar. Continuous transitions are denoted by a broken bar.
Unit cells of the corresponding stable phases are shown, where the big
(small) have a radius $R_A$ ($R_B$). }  
\label{fig:phase_diag_interface}
\end{figure}

\begin{figure}
\begin{center}
     \includegraphics[width=8.0cm]{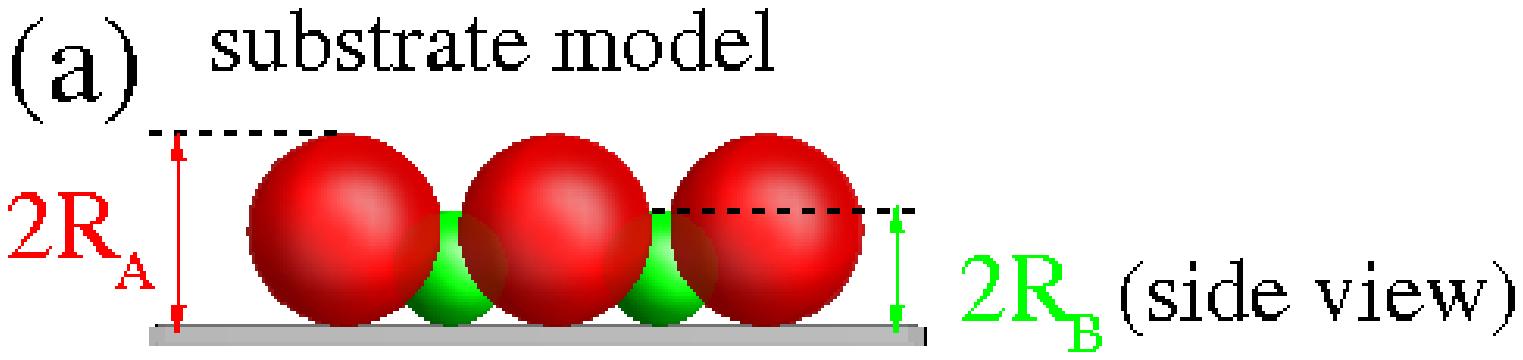}
     \includegraphics[width=8.0cm]{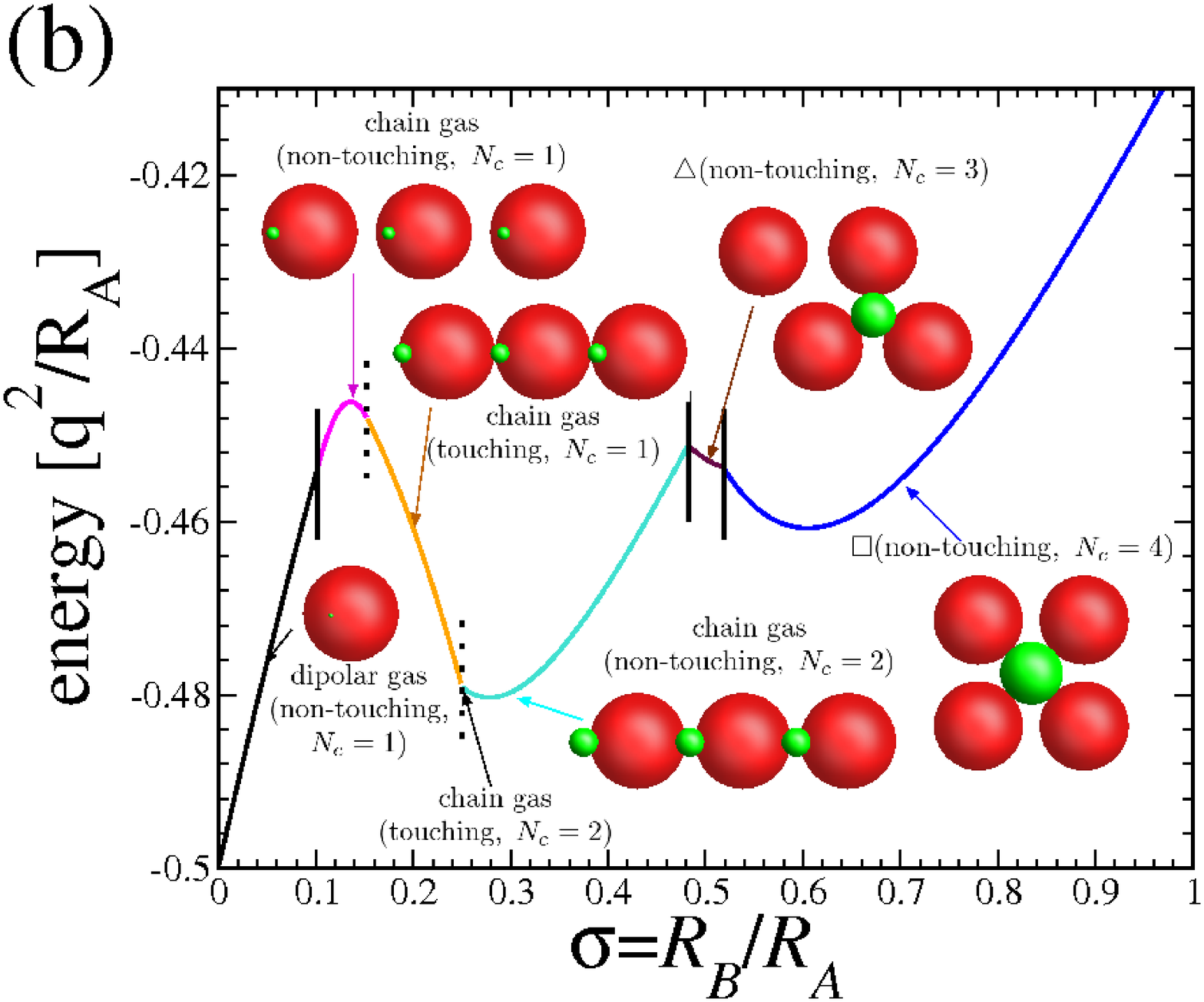}
\end{center}
\caption{Stable structures of oppositely charged spheres 
versus their size asymmetry $\sigma=R_B/R_A$ in the substrate model,
where all sphere surfaces touch the same plane: a)  side-view,
b)  (scaled) energy per ion. Discontinuous transitions 
between the structures are indicated by a solid bar. Continuous transitions are denoted by a broken bar.
Bottom views of the unit cells of the corresponding stable phases are displayed, where the big
(small) have a radius $R_A$ ($R_B$).} 
\label{fig:phase_diag_granular}
\end{figure}

We now consider two different set-ups, the ``interfacial model'' and the ``substrate model''.
In the interfacial model which can be considered as a purely two-dimensional situation,
the centres of all spheres are confined to a plane, see Fig. \ref{fig:phase_diag_interface}(a)
for a side view of a configuration. In the substrate model, on the other hand, all spheres
are touching the same underlying plane, see Fig. \ref{fig:phase_diag_granular}(a)
for a side view of a configuration.

For the interfacial model, the stability phase diagram  is shown 
versus the size asymmetry $\sigma$
in Fig. \ref{fig:phase_diag_interface}(b).
By increasing  $\sigma$, the following phase cascade occurs:
\begin{eqnarray*}
\triangle {\rm (touching},\ N_c=2{\rm )}  \to
\triangle {\rm (touching},\ N_c=3{\rm )}  \to \\  
\triangle {\rm (non{\mbox -}touching},\ N_c=3{\rm )} \to  
Rh {\rm (touching},\ N_c=3{\rm)} \to \\
\square {\rm (touching},\ N_c=4{\rm)} \to   
\square {\rm (non{\mbox-}touching},\ N_c=4{\rm)}   
\end{eqnarray*}
where the symbols $\triangle$ and $\square$ stand for triangular and square unit cells 
of the big ions $A$, respectively,
and  $Rh$ corresponds to rhombic unit cells, as  illustrated by the 
top views of the crystalline structures in Fig. \ref{fig:phase_diag_interface}(b). 
The topological aspect of these four phases can be discussed in terms of 
contact between the large spheres. 
More specifically, a ``touching'' configuration
involves connected big spheres while a ``non-touching'' one implies disconnected big spheres,
see Fig. \ref{fig:phase_diag_interface}(b).
In particular, 
the ``touching'' triangular and rhombic 
phases are both characterized by connected $A$-spheres in contrast to the ``non-touching''
$\triangle$ and $\square$ structures which possess disconnected large spheres.
Additionally, the ionic coordination number $N_c$ defined by the number of $A$ 
particles touching a single $B$ particle, is another relevant  topological 
characteristic.  

As far as the phase transitions reported in Fig. \ref{fig:phase_diag_interface}
are concerned, the following scenario takes place. 
At vanishing small ion size ($\sigma \to 0$), the small ions get squeezed
between two big ions so that the three centers of mass
lie on a same line. 
The first transition 
$\triangle {\rm (touching},\ N_c=2{\rm )}  \to 
\triangle {\rm (non{\mbox -}touching},\ N_c=3{\rm )}$
via the special structure $\triangle {\rm (touching},\ N_c=3{\rm )}$, 
characterized by an increase of the number of contacts between $A$ and
$B$ ions, occurs at $\sigma=\frac{2}{\sqrt 3} - 1$.
This special point corresponds to a compact triangular structure 
where a small ion has three contacts with neighbouring big ions, allowing a 
continuous transition.  
The second transition,    
$\triangle {\rm (non{\mbox -}touching},\ N_c=3{\rm )} \to  
Rh {\rm (touching},\ N_c=3{\rm)}$, occurring
at $\sigma=0.297$ is discontinuous as signaled by a jump of the 
angle between the two adjacent sides of the unit cell.  
The third transition  
$Rh {\rm (touching},\ N_c=3{\rm)}\to 
\square {\rm (non{\mbox-}touching},\ N_c=4{\rm)}$ via 
$\square {\rm (touching},\ N_c=4{\rm)}$, 
occurring at $\sigma=\sqrt 2 - 1$ is continuous.
A remarkable feature,  en passant, is the stability of the square phase over a wide range 
of the size ratio $\sigma$.

For the two non-touching  phases ($\triangle, \square$) one can assign two 
new Madelung constants $M$ associated to the (lattice) electrostatic energy per molecule
(i.e., a pair of ions $A$ and $B$) 
$
E = -M \frac{q^2}{R_A+R_B}.
$
Lekner sums carried on the appropriate lattices provide: 
$$
M_{\triangle} = 1.542,
\quad
M_{\square} = 1.616
$$
for the non-touching  triangular and  square structures, respectively. 
As expected, this reported value of $M_{\square}$ lies perfectly between that of the one dimensional 
lattice $(M_{1D}=2\ln 2)$ and that of a three-dimensional one ($M_{NaCl}=1.747$).

\begin{figure}
\begin{center}
\includegraphics[width=8.0cm]{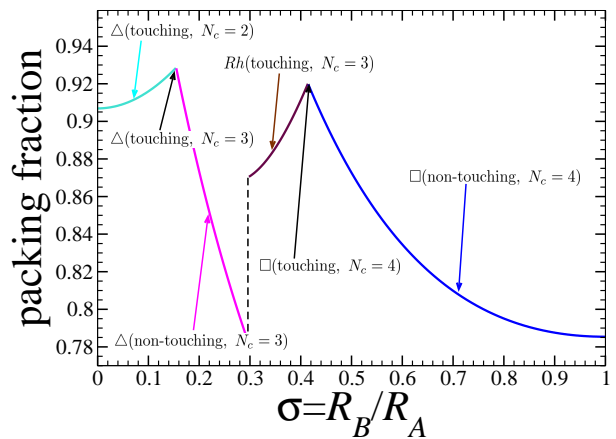}
\end{center}
\caption{
Area fraction as a function 
of the size aspect ratio $\sigma=R_B/R_A$.} 
\label{fig:packing_interface}
\end{figure}

Next we focus on the area fraction covered by the spheres which is defined as 
$\phi = \frac{\pi(R_A^2 + R_B^2)}{S_{cell}}$,
with $S_{cell}$ being the (projected) surface of the unit cell. The results are  
sketched in Fig. \ref{fig:packing_interface}. 
In the regime $\sigma < \frac{2}{\sqrt 3} - 1$ corresponding to compact triangular structures, 
there is enough space to host a small ion in the interstice offered by the touching big ions, 
so that the profile is identical to that of pure hard disk systems, where
$\phi=\frac{\pi(1+{\sigma}^2)}{2\sqrt 3}$.
In other words the location of the small sphere within the interstice
does not alter the packing fraction.
For non-touching triangular structures 
$(\frac{2}{\sqrt 3} - 1 < \sigma < 0.297)$, 
$\phi$ varies like  $\frac{2\pi(1+{\sigma}^2)}{3\sqrt 3(1+\sigma)^2}$.
In the rhombic phase regime $(0.297 < \sigma < \sqrt 2 - 1)$, 
the shape  as well as the surface of the unit cell
vary in a non-trivial manner.
For large enough small ions, in the square phase regime, 
$(\sqrt 2 - 1 < \sigma < 1)$, $\phi$ is  given by 
$\frac{\pi(1+{\sigma}^2)}{2 (1+\sigma)^2}$. 

We finally present results for the ``substrate model'' where all spheres are 
touching the same underlying substrate plane, see Fig. \ref{fig:phase_diag_granular}(a)
for a side-view. The stable crystalline structure and their energy per ion
is shown in Fig. \ref{fig:phase_diag_granular}(b) versus the diameter ratio $\sigma$.
For increasing size asymmetry $\sigma$, a cascade of six structures is found:
\begin{eqnarray*}
{\rm dipolar\ gas\  (non\mbox{-}touching},\ N_c=1{\rm )} \to\\
{\rm  chain\ gas\  (non\mbox{-}touching},\ N_c=1{\rm )} \to\\
{\rm  chain\ gas\  (touching},\ N_c=1{\rm )} \to\\
{\rm  chain\ gas\  (touching},\ N_c=2{\rm )} \to\\
{\rm  chain\ gas\  (non\mbox{-}touching },\ N_c=2 {\rm )} \to\\
\triangle  {\rm (non\mbox{-}touching},\ N_c=3{\rm )} \to 
\square  {\rm  (non\mbox{-}touching},\ N_c=4{\rm )}
\end{eqnarray*}
In the limit of very large asymmetry $(\sigma \to 0)$, dipoles perpendicular to the substrate 
plane are formed. Such parallel dipoles repel eachother and  arrange into a crystal with 
diverging lattice constant and a coordination number $N_c=1$. We call this state a ``{\it dipole gas}''.
When $\sigma$ increases, the dipole moments are getting gradually tilted relative to the substrate normal
until a first-order  transition towards a chain composed of dipolar pairs occurs
where the individual dipoles are non-touching and the coordination numbers stays at $N_c=1$.
The distance between neighbouring chains diverges since they are mutually repulsive.
We are dealing therefore with a ``{\it chain gas}'' at infinite dilution, i.e.\
the system is periodic in the direction along the string but with a diverging interchain 
distance perpendicular to it
bearing some analogy to smectic dipolar sheets \cite{Eisenmann}. Upon increasing $\sigma$ more,
the chain structure changes continuously  to an internal conformation with touching large spheres
(still with $N_c=1$). This structure then transforms continuously  into a chain gas of non$-$touching large spheres
with $N_c=2$. 
Then, there are two crystalline structures appearing known already 
from the interface model, namely a non-touching $\triangle$ and a non-touching $\square$
lattice. The former has only a tiny stability domain while the
latter is stable along an enormous range of  $0.517 < \sigma \leq 1$.
We emphasize the striking emergence of the gas phases in the substrate model 
which are absent in the interface model and in three dimensions.

In conclusion, we have explored the stable ground state structures of two-dimensional ionic crystals for
an ``interface'' and a ``substrate'' set-up at zero pressure. For a 1:1 oppositely charged mixture of spheres 
with different diameters, various stable crystalline phases were identified including
periodic crystals and chain and dipolar gases. Apart from adsorbed molecular salt systems,
the structures of the interface set-up can be verified
in suspensions of oppositely charged particles. The substrate set-up, on the other hand, is realized for
oppositely charged granular matter on a plane. 
Recent experiments on oppositely charged granular sheets with $\sigma\approx 1$ \cite{Whitesides1,Whitesides2}  have indeed
revealed a stable $\square {\rm (non\mbox{-}touching)}$ configuration which is 
confirmed by our calculations.
More experimental investigations on systems with higher size asymmetry
are performable  and could open
the way to see the $\triangle {\rm (non\mbox{-}touching)}$ and the predicted chain and dipolar gases.

The penalty method can in principle be applied to any other potentials which involve 
hard-body parts, both in two and in three dimensions. It would be interesting to see the stability
phase diagram for different mixtures as e.g. colloidal
hard-spheres mixtures with or without added nonadsorbing polymers.

Financial support from DFG (SFB TR6, D1) is acknowledged.




\begin{thebibliography}{99}

\bibitem{Catlow} S. M. Woodley and R. Catlow, 
Nature Materials {\bf 7}, 937 (2008).


\bibitem{Madelung} E. Madelung, Phys. Z. {\bf 19}, 542 (1918).

\bibitem{Ewald} P. P. Ewald, Ann. Phys. {\bf 64}, 253 (1921).

\bibitem{Evans} See e.g.: R. C. Evans, {\it An Introduction to Crystal Chemistry}, 
Cambridge University Press, pages 41-43, 1966.

\bibitem{Leunissen} M. E Leunissen, C. G. Christova, A.P Hynninen,C. P Royall, A. I. Campbell, A. Imhof, M. Dijkstra, R. van Roij, A. van Blaaderen, Nature {\bf 437}, 235 (2005).  

\bibitem{Dijkstra} A. P. Hynninen, C. G. Christova, R. van Roij, A. van Blaaderen, M. Dijkstra
 Phy. Rev. Lett.  {\bf 96}, 138308  (2006).

\bibitem{Dijkstra2} A. P. Hynninen, M. E. Leunissen, A. van Blaaderen {\it et al.}
Phys. Rev. Lett. {\bf 96}, 018303 (2006).


\bibitem{Rosendahl} S. M. Rosendahl, I. J. Burgess, Electrochimica Acta {\bf 53},  6759 (2008).

\bibitem{traps}
M. C. Jenkins, S. U. Egelhaaf,
J. Phys.: Condens. Matter {\bf 20}, 404220 (2008).

\bibitem{interface_Goedel}
C. E. McNamee, M. Jaumann, M. Moller, A. L. Ding, S. Hemeltjen, S. Ebert, W. Baumann, W. A. Goedel,
Langmuir {\bf 21},  10475 (2005).

\bibitem{Whitesides1} G. K. Kaufman, S. W. Thomas, M. Reches, B. F. Shaw, J. Feng, G. M. Whitesides, 
Soft Matter   {\bf 5}, 1188 (2009).

\bibitem{Whitesides2} G. K. Kaufman, M. Reches, S. W. Thomas, B. F. Shaw, J. Feng, G. M. Whitesides,
App. Phys. Lett. {\bf 94}, 044102  (2009).   

\bibitem{Pine}
V. N. Manoharan, M. T. Elsesser, D. J. Pine,  
Science {\bf 301}, 483 (2003).

\bibitem{Bechinger_PRL} J. Baumgartl, M. Zvyagolskaya, C. Bechinger, 
Phys. Rev. Lett. {\bf 99}, 205503  (2007).

\bibitem{Kecht} J. Kecht, B. Mihailova, K. Karaghiosoff, S. Mintova, T. Bein, 
Langmuir {\bf 20}, 5271 (2004).

\bibitem{Goedel1} For uncharged particles, this was realized in: F. Yan and W. A. Goedel,
Chem. Mater. {\bf 16}, 1622 (2004); Nano Lett. {\bf 4}, 1193 (2004).


\bibitem{nanowires}  K.-S. Cho, D. V. Talapin, W. Gaschler, 
C. B. Murray, J. Am. Chem. Soc. {\bf 127}, 7140 (2005).

\bibitem{Nagayama} K. Nagayama, S. Takeda, S. Endo, H. Yoshimura, 
Jpn. J. Appl. {\bf 34}, 3947 (1995). 

\bibitem{Fisher_Levin} Y. Levin and M. E. Fisher, 
Physica A {\bf 225}, 164 (1996).

\bibitem{Caillol_1998} J. J. Weis, D. Levesque, J. M. Caillol,
J. Chem. Phys. {\bf 109}, 7486 (1998).
\bibitem{GA}  D. Gottwald, C. N. Likos, G. Kahl, H. L{\"o}wen, 
 Phys. Rev. Letters {\bf 92}, 068301  (2004).

\bibitem{PRE_Dijkstra_2009} L. Filion and M. Dijkstra, 
Phys. Rev. E {\bf 79}, 046714 (2009).

\bibitem{Fiacco} A. V. Fiacco and G. P. McCormick, {\it Nonlinear Programming: Sequential  Unconstrained Minimization Techniques}, 
John Wiley and Sons, New York, 1968.

\bibitem{simplex} J.A. Nelder and R. Mead, {\it A simplex method for function minimization}, Computer Journal {\bf 7}, 308 (1965).



\bibitem{brodka_lekner} A. Grzybowski and A. Brodka, Mol. Phys. {\bf 100}, 1017 (2002).

\bibitem{Eisenmann} C. Eisenmann, U. Gasser, P. Keim, G. Maret, Phys. Rev. Letters {\bf 93}, 
 105702  (2004).


\end{thebibliography}
\end{document}